%Paper: gr-qc/9403006
%From: rwt@lvu.ph.lancs.ac.uk (Robin W Tucker)
%Date: Wed, 2 Mar 94 9:42:17 WET

%qgxycqg.tex

% The following settings are for Plain TeX:

%%%%%%%%%%%%%%%%%%%%%%%%%%%%%%%%%%%%%%%
%% begin Tevian's twelvepoint macros %%
%%%%%%%%%%%%%%%%%%%%%%%%%%%%%%%%%%%%%%%

%% Version 1.0 of 9/21/91.
%% (twelve point footline added 12/13/91)
%% Based on macros originally obtained from IAS, Princeton.

%% The only things which should need to be adjusted for most standard
%% installations are the title fonts and the AMS fonts; see below.

%% Adjust margins for A4 paper.
%
% \hoffset=4pt
% \vsize=9.8in
% \voffset=-0.3in

%% Define standard fonts
%
 \font\twelvebf=cmbx12
 \font\twelvett=cmtt12
 \font\twelveit=cmti12
 \font\twelvesl=cmsl12
 \font\twelverm=cmr12		\font\ninerm=cmr9
 \font\twelvei=cmmi12		\font\ninei=cmmi9
 \font\twelvesy=cmsy10 at 12pt	\font\ninesy=cmsy9
 \skewchar\twelvei='177		\skewchar\ninei='177
 \skewchar\seveni='177	 	\skewchar\fivei='177
 \skewchar\twelvesy='60		\skewchar\ninesy='60
 \skewchar\sevensy='60		\skewchar\fivesy='60
%
%% The following are built into Plain TeX.
%
%\font\sevenrm=cmr7		\font\fiverm=cmr5
%\font\seveni=cmmi7		\font\fivei=cmmi5
%\font\sevensy=cmsy7		\font\fivesy=cmsy5
%\font\tenex=cmex10
%\font\tenrm=cmr10	\font\teni=cmmi10	\font\tensy=cmsy10
%\font\tenbf=cmbx10	\font\tentt=cmtt10
%\font\tenit=cmti10	\font\tensl=cmsl10

%% Define large fonts for titles.
%
 \font\fourteenrm=cmr12 scaled 1200
 \font\seventeenrm=cmr12 scaled 1440
 \font\fourteenbf=cmbx12 scaled 1200
 \font\seventeenbf=cmbx12 scaled 1440
%
%% Alternative definition if the above fonts are missing.
%
%\font\fourteenrm=cmr10 scaled 1440
%\font\seventeenrm=cmr10 scaled 1728
%\font\fourteenbf=cmbx10 scaled 1440
%\font\seventeenbf=cmbx10 scaled 1728

%% Now define the (old) AMS fonts needed for blackboard bold.
%
%\font\tenmsy=msym10
%\font\twelvemsy=msym10 scaled 1200
%\newfam\msyfam
%
%% Alternatively, here are the new AMS fonts for blackboard bold.
%
\font\tenmsb=msbm10
\font\twelvemsb=msbm10 scaled 1200
\newfam\msbfam

%% Define fonts for small caps.
%
\font\tensc=cmcsc10
\font\twelvesc=cmcsc10 scaled 1200
\newfam\scfam

%% Define macros for different sizes.
%
\def\seventeenpt{\def\rm{\fam0\seventeenrm}%
 \textfont\bffam=\seventeenbf	\def\bf{\fam\bffam\seventeenbf}}
\def\fourteenpt{\def\rm{\fam0\fourteenrm}%
 \textfont\bffam=\fourteenbf	\def\bf{\fam\bffam\fourteenbf}}
\def\twelvept{\def\rm{\fam0\twelverm}%
 \textfont0=\twelverm	\scriptfont0=\ninerm	\scriptscriptfont0=\sevenrm
 \textfont1=\twelvei	\scriptfont1=\ninei	\scriptscriptfont1=\seveni
 \textfont2=\twelvesy	\scriptfont2=\ninesy	\scriptscriptfont2=\sevensy
 \textfont3=\tenex	\scriptfont3=\tenex	\scriptscriptfont3=\tenex
 \textfont\itfam=\twelveit	\def\it{\fam\itfam\twelveit}%
 \textfont\slfam=\twelvesl	\def\sl{\fam\slfam\twelvesl}%
 \textfont\ttfam=\twelvett	\def\tt{\fam\ttfam\twelvett}%
 \scriptfont\bffam=\tenbf 	\scriptscriptfont\bffam=\sevenbf
 \textfont\bffam=\twelvebf	\def\bf{\fam\bffam\twelvebf}%
 \textfont\scfam=\twelvesc	\def\sc{\fam\scfam\twelvesc}%
%\textfont\msyfam=\twelvemsy	%  Old AMS font
 \textfont\msbfam=\twelvemsb	%  New AMS font
 \baselineskip 14pt%
 \abovedisplayskip 7pt plus 3pt minus 1pt%
 \belowdisplayskip 7pt plus 3pt minus 1pt%
 \abovedisplayshortskip 0pt plus 3pt%
 \belowdisplayshortskip 4pt plus 3pt minus 1pt%
 \parskip 3pt plus 1.5pt
 \setbox\strutbox=\hbox{\vrule height 10pt depth 4pt width 0pt}}
\def\tenpt{\def\rm{\fam0\tenrm}%
 \textfont0=\tenrm	\scriptfont0=\sevenrm	\scriptscriptfont0=\fiverm
 \textfont1=\teni	\scriptfont1=\seveni	\scriptscriptfont1=\fivei
 \textfont2=\tensy	\scriptfont2=\sevensy	\scriptscriptfont2=\fivesy
 \textfont3=\tenex	\scriptfont3=\tenex	\scriptscriptfont3=\tenex
 \textfont\itfam=\tenit		\def\it{\fam\itfam\tenit}%
 \textfont\slfam=\tensl		\def\sl{\fam\slfam\tensl}%
 \textfont\ttfam=\tentt		\def\tt{\fam\ttfam\tentt}%
 \scriptfont\bffam=\sevenbf 	\scriptscriptfont\bffam=\fivebf
 \textfont\bffam=\tenbf		\def\bf{\fam\bffam\tenbf}%
 \textfont\scfam=\tensc		\def\sc{\fam\scfam\tensc}%
%\textfont\msyfam=\tenmsy	%  Old AMS font
 \textfont\msbfam=\tenmsb	%  New AMS font
 \baselineskip 12pt%
 \abovedisplayskip 6pt plus 3pt minus 1pt%
 \belowdisplayskip 6pt plus 3pt minus 1pt%
 \abovedisplayshortskip 0pt plus 3pt%
 \belowdisplayshortskip 4pt plus 3pt minus 1pt%
 \parskip 2pt plus 1pt
 \setbox\strutbox=\hbox{\vrule height 8.5pt depth 3.5pt width 0pt}}

%% Define size switching macros.
%
\def\twelvepoint{%
 \def\small{\tenpt\rm}%
 \def\normal{\twelvept\rm}%
 \def\big{\fourteenpt\rm}%
 \def\huge{\seventeenpt\rm}%
 \footline{\hss\twelverm\folio\hss}
 \normal}
%

%% The following are for backwards compatibility with earlier versions.
%
\def\bigbold{\big\bf}

%% Define general macros.
%
\catcode`\@=11
%
%% Redefine footnotes to use a smaller font.
%
\def\footnote#1{\edef\@sf{\spacefactor\the\spacefactor}#1\@sf
 \insert\footins\bgroup\small
 \interlinepenalty100	\let\par=\endgraf
 \leftskip=0pt		\rightskip=0pt
 \splittopskip=10pt plus 1pt minus 1pt	\floatingpenalty=20000
 \smallskip\item{#1}\bgroup\strut\aftergroup\@foot\let\next}
%
%% Define Blackboard bold using old AMS fonts.
%
%\def\Bbb{\ifmmode\let\next\Bbb@\else
% \def\next{\errmessage{Use \string\Bbb\space only in math mode}}\fi\next}
%\def\Bbb@#1{{\Bbb@@{#1}}}
%\def\Bbb@@#1{\fam\msyfam#1}
%
%% Define Blackboard bold using new AMS fonts.
%% (Font definitions and reference in tenpt and twelvept must be enabled.)
%
\def\hexnumber@#1{\ifcase#1 0\or 1\or 2\or 3\or 4\or 5\or 6\or 7\or 8\or
 9\or A\or B\or C\or D\or E\or F\fi}
\edef\msbfam@{\hexnumber@\msbfam}

%
%% If you don't have any AMS fonts, comment out both sets of macros above and
%% use the following line instead.
%
%\def\Bbb#1{{\cal #1}}
%
\catcode`\@=12
%% These macros are designed to be used with Plain TeX to automatically number
%% equations, footnotes, figures, and references.
%% (However, unlike LaTeX, no auxiliary files are used, so that the source
%% file only needs to be TeX'd once!)

\newcount\EQNO      \EQNO=0
\newcount\FIGNO     \FIGNO=0
\newcount\REFNO     \REFNO=0
\newcount\SECNO     \SECNO=0
\newcount\SUBSECNO  \SUBSECNO=0
\newcount\FOOTNO    \FOOTNO=0
\newbox\FIGBOX      \setbox\FIGBOX=\vbox{}
\newbox\REFBOX      \setbox\REFBOX=\vbox{}
\newbox\RefBoxOne   \setbox\RefBoxOne=\vbox{}
\expandafter\ifx\csname normal\endcsname\relax\def\normal{\null}\fi

\def\Eqno{\global\advance\EQNO by 1 \eqno(\the\EQNO)%
    \gdef\label##1{\xdef##1{\nobreak(\the\EQNO)}}}
\def\Fig#1{\global\advance\FIGNO by 1 Figure~\the\FIGNO%
    \global\setbox\FIGBOX=\vbox{\unvcopy\FIGBOX
      \narrower\smallskip\item{\bf Figure \the\FIGNO~~}#1}}
\def\Ref#1{\global\advance\REFNO by 1 \nobreak[\the\REFNO]%
    \global\setbox\REFBOX=\vbox{\unvcopy\REFBOX\normal
      \smallskip\item{\the\REFNO .~}#1}%
    \gdef\label##1{\xdef##1{\nobreak[\the\REFNO]}}}
\def\Section#1{\SUBSECNO=0\advance\SECNO by 1
    \bigskip\leftline{\bf \the\SECNO .\ #1}\nobreak}
\def\Subsection#1{\advance\SUBSECNO by 1
    \medskip\leftline{\bf \ifcase\SUBSECNO\or
    a\or b\or c\or d\or e\or f\or g\or h\or i\or j\or k\or l\or m\or n\fi
    )\ #1}\nobreak}
\def\Footnote#1{\global\advance\FOOTNO by 1
    \footnote{\nobreak$\>\!{}^{\the\FOOTNO}\>\!$}{#1}}
\def\SameFootnote{$\>\!{}^{\the\FOOTNO}\>\!$}

\def\References{\bigskip\centerline{\bf REFERENCES}
                \smallskip\copy\REFBOX}
\def\NewRefPage{\setbox\RefBoxOne=\vbox{\unvcopy\REFBOX}
		\setbox\REFBOX=\vbox{}
		\def\References{\bigskip\centerline{\bf REFERENCES}
                		\nobreak\smallskip\nobreak\copy\RefBoxOne
				\vfill\eject
				\smallskip\copy\REFBOX}
		\def\NewRefPage{}}

% bmit.tex

\font\twelvebm=cmmib10 at 12pt
\font\tenbm=cmmib10
\font\ninei=cmmi9
\newfam\bmfam

\def\twelvepointbmit{
\textfont\bmfam=\twelvebm
\scriptfont\bmfam=\ninei
\scriptscriptfont\bmfam=\seveni
\def\bmit{\fam\bmfam\twelvebm}
}

\def\tenpointbmit{
\textfont\bmfam=\tenbm
\scriptfont\bmfam=\seveni
\scriptscriptfont\bmfam=\fivei
\def\bmit{\fam\bmfam\tenbm}
}

\tenpointbmit

\mathchardef\Gamma="7100
\mathchardef\Delta="7101
\mathchardef\Theta="7102
\mathchardef\Lambda="7103
\mathchardef\Xi="7104
\mathchardef\Pi="7105
\mathchardef\Sigma="7106
\mathchardef\Upsilon="7107
\mathchardef\Phi="7108
\mathchardef\Psi="7109
\mathchardef\Omega="710A
\mathchardef\alpha="710B
\mathchardef\beta="710C
\mathchardef\gamma="710D
\mathchardef\delta="710E
\mathchardef\epsilon="710F
\mathchardef\zeta="7110
\mathchardef\eta="7111
\mathchardef\theta="7112
\mathchardef\iota="7113
\mathchardef\kappa="7114
\mathchardef\lambda="7115
\mathchardef\mu="7116
\mathchardef\nu="7117
\mathchardef\xi="7118
\mathchardef\pi="7119
\mathchardef\rho="711A
\mathchardef\sigma="711B
\mathchardef\tau="711C
\mathchardef\upsilon="711D
\mathchardef\phi="711E
\mathchardef\cho="711F
\mathchardef\psi="7120
\mathchardef\omega="7121
\mathchardef\varepsilon="7122
\mathchardef\vartheta="7123
\mathchardef\varpi="7124
\mathchardef\varrho="7125
\mathchardef\varsigma="7126
\mathchardef\varphi="7127

%%%%%%%%%%%%%%%%%%%%%%%%%%%%%%%%%%%%%%%%%%%%%%%%%%%%%%%%%%%%%%%%%%%%%%%%%%%%%

\twelvepoint			% Turn on 12 pt fonts.

\font\twelvebm=cmmib10 at 12pt
\font\tenbm=cmmib10
\font\ninei=cmmi9
\newfam\bmfam

\def\twelvepointbmit{
\textfont\bmfam=\twelvebm
\scriptfont\bmfam=\ninei
\scriptscriptfont\bmfam=\seveni
\def\bmit{\fam\bmfam\twelvebm}
}

\def\tenpointbmit{
\textfont\bmfam=\tenbm
\scriptfont\bmfam=\seveni
\scriptscriptfont\bmfam=\fivei
\def\bmit{\fam\bmfam\tenbm}
}

\tenpointbmit

\mathchardef\Gamma="7100
\mathchardef\Delta="7101
\mathchardef\Theta="7102
\mathchardef\Lambda="7103
\mathchardef\Xi="7104
\mathchardef\Pi="7105
\mathchardef\Sigma="7106
\mathchardef\Upsilon="7107
\mathchardef\Phi="7108
\mathchardef\Psi="7109
\mathchardef\Omega="710A
\mathchardef\alpha="710B
\mathchardef\beta="710C
\mathchardef\gamma="710D
\mathchardef\delta="710E
\mathchardef\epsilon="710F
\mathchardef\zeta="7110
\mathchardef\eta="7111
\mathchardef\theta="7112
\mathchardef\iota="7113
\mathchardef\kappa="7114
\mathchardef\lambda="7115
\mathchardef\mu="7116
\mathchardef\nu="7117
\mathchardef\xi="7118
\mathchardef\pi="7119
\mathchardef\rho="711A
\mathchardef\sigma="711B
\mathchardef\tau="711C
\mathchardef\upsilon="711D
\mathchardef\phi="711E
\mathchardef\cho="711F
\mathchardef\psi="7120
\mathchardef\omega="7121
\mathchardef\varepsilon="7122
\mathchardef\vartheta="7123
\mathchardef\varpi="7124
\mathchardef\varrho="7125
\mathchardef\varsigma="7126
\mathchardef\varphi="7127

%>>>>>>>>>>>>>>>>>>>>>>>>>>>>>>>>>>>>>>>>>>>>>>>>>>>>>>>>>>>>>>>>>>>>>>>

%%%%%%%%%%%%%%%%%%%%%%%%%%%%%%%%%%%%%%%%%%%%%%%%%%%%%%%%%%%%%%%%%%%%%%%%%%%%

\relax
%

%%%%%%%%%%%%%%%%%%%%%%%%%%%%%%%%%%%%%%%%%%%%%%%%%%%%%%%%%%%%%%%%%%%%%%%%%%

\twelvepointbmit		% Turn on 12 pt bold math italics.

\def\sqr#1#2{{\vbox{\hrule height.#2pt
 	\hbox{\vrule width.#2pt height#1pt \kern#1pt\vrule width.#2pt}
		\hrule height.#2pt}}}

\def\Partial#1{\partial_{#1}^{\raise2pt\hbox{$\scriptstyle 2$}}}
\def\bar{\overline}

\def\Uin{u^{\,\raise2pt\hbox{$\scriptstyle\rm in$}}}
\def\Uout{u^{\,\raise2pt\hbox{$\scriptstyle\rm out$}}}

%**end of header

\nopagenumbers

\def\today{\number\day\space\ifcase\month\or
  January\or February\or March\or April\or May\or June\or
  July\or August\or September\or October\or November\or December\fi
  \space\number\year}
%\rightline{DRAFT,\space\today}
%\rightline{ qgxy.tex }
\bigskip\bigskip

\null\bigskip\bigskip\bigskip

\baselineskip=27pt

%%%%%%%%%%%%%%%%%%%%%%%%%%%%%%%%%%%%%%%%%%%%%%%%%%%%%%%%%%%%%%%%%%%%%%%%%%
% START
\vskip 1cm

%%%%%%%%%%%%%%%%%%%%%%%%%%%%%%%%%%%%%%%%%%%%%%%%%%%%%%%%%%%%%%%%%%%%%%%%%%%%
%\magnification=\magstep1       	% Imitate 12 pt output from Plain TeX.
%\font\bigbold=cmbx10 scaled 1200
%\def\Bbb#1{{\cal #1}}          	% Imitate blackboard bold.

% The following settings use Tevian's twelvepoint macros:
%
\input tevians_macros		% Robin's version of tw and number
\twelvepoint			% Turn on 12 pt fonts.
\input bmit			% Tevian's macros for bold math italics.
\twelvepointbmit		% Turn on 12 pt bold math italics.

\def\sqr#1#2{{\vbox{\hrule height.#2pt
 	\hbox{\vrule width.#2pt height#1pt \kern#1pt\vrule width.#2pt}
		\hrule height.#2pt}}}

\def\Partial#1{\partial_{#1}^{\raise2pt\hbox{$\scriptstyle 2$}}}
\def\bar{\overline}

\def\Uin{u^{\,\raise2pt\hbox{$\scriptstyle\rm in$}}}
\def\Uout{u^{\,\raise2pt\hbox{$\scriptstyle\rm out$}}}

%**end of header

\nopagenumbers

\def\today{\number\day\space\ifcase\month\or
  January\or February\or March\or April\or May\or June\or
  July\or August\or September\or October\or November\or December\fi
  \space\number\year}
%\rightline{DRAFT,\space\today}
%\rightline{ qgxy.tex }
\bigskip\bigskip

\null\bigskip\bigskip\bigskip
\centerline{\bigbold ON THE RELATION BETWEEN}
\centerline{\bigbold CLASSICAL AND QUANTUM COSMOLOGY}
\centerline{\bigbold IN A  2-DIMENSIONAL DILATON-GRAVITY MODEL}
\bigskip\bigskip\bigskip

\centerline{M \"Onder${}^{\dagger}$}
\medskip
\centerline{Robin W. Tucker}
\medskip
\medskip
\centerline{\it School of Physics and Materials,}
\centerline{\it University of Lancaster,
		Bailrigg, Lancs. LA1 4YB, UK}
\centerline{\tt rwt{\rm @}lavu.physics.lancaster.ac.uk}

\bigskip\bigskip\bigskip\bigskip
\centerline{\bf ABSTRACT}
\midinsert
\narrower\narrower\noindent

We analyse the classical and quantum theory of a scalar field
interacting with gravitation in two dimensions. We describe a class
 of analytic solutions to the Wheeler-DeWitt equation
from which we are able
to synthesise states that give prominence to a set of classical cosmologies.
These states  relate in a remarkable way to the general solution of the
classical field equations. We express these relations, without
approximation, in terms of a metric and a closed form on the domain of
quantum states.

\endinsert

\vfill

${}^{\dagger}$
{\it Department of Physics Engineering, Hacettepe University, 06532
 Beytepe, Ankara, \break  Turkey.} {\tt F\_Onder{\rm @}trhun}

\eject

\headline={\hss\rm -~\folio~- \hss}     % Adds page numbers

\def\pprime{^\prime}
\def\frac#1#2{{#1\over #2}}

\Section{Introduction}

The application of quantum theory to the Universe requires a re-orientation
of a number of concepts that are traditionally invoked in order to
interpret the theory in terms of a pre-existing classical spacetime. With
the description of classical gravitation
 and matter in terms of a spacetime geometry
it becomes necessary to assign  quantum amplitudes to matter and
space-geometry configurations and to attempt a description of
 classical spacetimes
in terms of particular enhancements  of  such amplitudes. Among the many
problems
that need to be addressed in this approach are the consistent quantisation
of gravity and matter and the subsequent emergence of a time-oriented
classical spacetime. Since the pioneering efforts of
\Ref{ B S DeWitt, Phys. Rev. {\bf 160}  (1967) 1113}, \Ref{P A M Dirac,
Can. J. Math. {\bf 2} (1950) 129; Proc. Roy. Soc. (London) {\bf A246}
(1958) 326; ibid 333},
\Ref{C W Misner, Phys. Rev. {\bf 186} (1969) 1319}
\Ref{W F Blyth, C J Isham, Phys. Rev. {\bf D11} (1975) 768}
\Ref{M A H MacCallum, in ``Quantum Gravity: an Oxford Symposium'', eds. C J
Isham, R Penrose and D W Sciama, 1975},
\Ref{D Brill, R H Gowdy, Rep. Prog. Phys. {\bf 33} (1970) 413}
\Ref{ J B Hartle, S W Hawking, Phys.Rev. {\bf D28} (1983) 2960}
the subject has undergone a considerable revival of interest.
Part of this revival has come from studies of 2-D models which have arisen
either from string-inspired limits or the suppression  of inhomogeneous
modes in Einstein's theory of general relativity,
\Ref{G Mandal, A M Sengupta, S R Wadia, Mod.Phys. Lett. {\bf A6}  (1991)
1685},
\Ref{E Witten, Phys. Rev. {\bf D44}  (1991) 314 }\label\witten,
\Ref{C G Callan, Jr., S B Giddings, J A Harvey, A Strominger, Phys. Rev. {\bf
D45}  (1992) R1005 },
\Ref{S Chaudhuri, D Minic, ``On the Black Hole Background of
Two-Dimensional String Theory'', Austin Preprint UTTG-30-92 (1992) },
 \Ref{J Navarro-Salas, M Navarro,V Aldaya,
`` Wave Functions of the Induced 2D-Gravity'',
 Valencia Preprint FTUV/93-3}\label \ald ,
\Ref{ R Jackiw, `` Gauge Theories for Gravity on a Line'', ({\it In Memoriam} M
C
Polivanov), MIT Preprint 1993}.

In \Ref{T Dereli, R W Tucker, Class. Quantum Grav.
{\bf 10} (1993) 365} it was observed that by formulating Einstein's
field equations on a\break non-Riemannian (torsion-free) space-time it was
possible to find
cosmological solutions with degenerate metrics describing transitions
between Lorentzian and Euclidean signatures.
 In \Ref{ T Dereli, M \"Onder, R W Tucker,
 Class. Quantum Grav.{\bf 10} (1993) 1425}, by choosing a
particular set of variables to quantise, we exhibited a family of Hilbert
spaces describing exact solutions of the Wheeler-DeWitt equation and
offered tentative evidence suggesting that it was possible to synthesise
non-dispersive wave packets that peaked in the vicinity of classical
minisuperspace submanifolds. Furthermore, by arguing that classical time
could be associated with a choice of parametrisation of such dynamically
prescribed submanifolds we were able to place classical signature changing
solutions within the context of quantum cosmology. In an attempt to make
more precise some of these ideas we have sought this phenomena in the
framework of a particular classically soluble 2-D model of gravitation.
The model consists of a scalar field $\psi$, representing matter, coupled
directly to the scalar curvature of a 2-D manifold and a potential of the
form
$U(\psi)=\Lambda_0+\alpha e^{c\psi}$. This interaction ensures that
gravity on the line is non-trivial and the choice of potential enables the
cosmological sector of the theory to be reduced to the study of a
constrained isotropic oscillator-ghost-oscillator system. In this model we
are able to address some of the problems above and in particular find
exact quantum amplitudes for classical signature changing configurations.
Thus  we construct
non-dispersive quantum states
that may be used to define classical
geometries, some of which possess degenerate non-Riemannian metrics.
These states, generated from particular solutions to the Wheeler-DeWitt
equation,  relate in a remarkable way to the general solution of the
classical field equations. We express these relations, without
approximation, in terms of a metric and a closed form on the domain of
quantum states.

\Section{The Model}

\def\vv{{ V}}
\def\nn{{\cal N}}

\def\pf{\frac{\pi}{4}}
\def\thm{\theta_1-\theta_2}
\def\thp{\theta_1+\theta_2}

\def\sqrtc{\sqrt{c}}
\def\rp#1{\frac{1}{#1}}
Our quantisation programme begins with the classical action
$$
S[g,\psi]=\int_N\,\,\left\{ \frac 12\psi\star{\cal R}+cd\psi\wedge\star
d\psi +\star U(\psi)\right\}\Eqno$$
where $N$ is some domain of a two-dimensional manifold, $\psi$ is a real
scalar field and ${\cal R}$ is the curvature scalar of the Levi-Civita
connection associated with the metric tensor $g$. $\star$ denotes the Hodge
map of $g$. The real parameter $c>0$ and the real potential $U(\psi)$ will
be correlated below. In terms of local coordinates $\{s,t\}$ the local
Lagrangian density is defined by
$$S=\int_N{\cal L}\,ds\wedge dt\Eqno$$
where
$${\cal L}=(\vert det(g)\vert )^{1/2}\,\,\left\{\frac 12 \psi{\cal
R}+c \,(\nabla\psi)^2 +U(\psi)\right\}.\Eqno$$
(For technical simplicity we shall assume that this space-time is globally
$\Sigma_t \times \Sigma_s$ where $\Sigma_s$ is compact.)
In  \Ref{M \"Onder, R W
Tucker,
%``Canonical Analysis of a Class of 2-Dimensional Dilaton-Gravity Models'',
 Phys. Letts. {\bf B311} (1993) 47} \label\rwt
 we developed the classical theory with a Lorentzian signature
metric
 and chose to parametrise the coordinate components of the metric in terms
of the lapse function $\nn$ and the shift function $\vv$:
$$g=-e^2\otimes e^2+e^1\otimes e^1.\Eqno$$
$$e^1=\frac 1\chi (ds-\vv dt)\Eqno$$
$$e^2=\nn dt\Eqno$$
Here we introduce the invertible mapping $(\psi,\chi)\mapsto (x,y)$:
$$x=\frac{1}{2\sqrtc}\left ( \frac{e^{c\psi}}{\chi}+e^{-c\psi}\right )
 \label\mapa\Eqno$$
$$y=\frac{1}{2\sqrtc}\left ( \frac{e^{c\psi}}{\chi}-e^{-c\psi}\right )
\Eqno \label\mapb$$
on a domain that maintains the reality of $\psi$.
We find in terms of $x$ and $y$
$${\cal L}(x,\dot x,x\pprime,y,\dot y,y\pprime)=
\frac{1}{\nn}\left\{\dot x^2-\dot y^2+2\vv(x\pprime\dot
x-y\pprime\dot y)+\vv\pprime(x+y)(\dot x-\dot y)\right\}+$$
$$\frac{\vv^2}{\nn}({x\pprime}^2-{y\pprime}^2)+
\frac{\vv\vv\pprime}{\nn}(x+y)(x\pprime-y\pprime)+
\nn\left\{\frac{1}{c^2}\frac{(x\pprime-y\pprime)^2}{(x^2-y^2)(x-y)^2}+c(x^2-y^2)U(x,y)\right\}-$$
$$\frac{\nn\pprime}{c^2}\frac{(x\pprime-y\pprime)}{(x^2-y^2)(x-y)}\Eqno$$
where $\dot x=\partial_t x,\, x\pprime=\partial_s x$ etc.
It follows that the Hamiltonian  may be written
$$H[x,p_x,y,p_y]=
\int_N (\nn{\Phi_1}+\vv{\Phi_2})\,ds\Eqno$$ in terms of the
 first class  classical constraints
$${\Phi_1}=\rp4 p_x^2-\rp4 p_y^2-c(x^2-y^2)U(x,y)-$$
$$\frac{1}{c^2}\frac{({x\pprime}\pprime-{y\pprime}\pprime)}{(x^2-y^2)(x-y)}+
\frac{2}{c^2}\frac{(x\pprime-y\pprime)(x x\pprime-y
y\pprime)}{(x^2-y^2)^2(x-y)}\Eqno$$
$${\Phi_2}=\frac{1}{2}\left\{(x+y)(p_x\pprime +p_y\pprime)-(x\pprime
p_x+y\pprime p_y)+(x\pprime p_y+y\pprime p_x)\right\}.\Eqno$$
There are no further constraints.
{}From the structure of these equations we are prompted to consider models
defined by a potential of the form:
$$U(\psi)=\Lambda_0+\alpha e^{c\psi}\Eqno \label\pot$$
where $\Lambda_0$ and $\alpha$ are real constants, since for cosmological
considerations where all fields are independent of $s$, and $\Lambda_0$ and
$\alpha$ are assumed non-zero, the Lagrangian
density takes a simple form in the gauge $\nn=1,\,\,\vv=0$ :
$${\cal L}=(\dot X^2-\dot Y^2)+\Lambda_0c(X^2-Y^2)\Eqno \label\lag$$
where
$$X=x+\frac{\alpha}{2\Lambda_0\sqrtc}\Eqno$$
$$Y=y-\frac{\alpha}{2\Lambda_0\sqrtc}.\Eqno$$
Furthermore the Hamiltonian constraint becomes
$${\Phi_1}=\rp4 (p_X^2-p_Y^2)-\Lambda_0c(X^2-Y^2).\Eqno$$
Thus if we take a negative $\Lambda_0$ and write
$\Lambda_0c=-\omega^2$
we map the theory onto a constrained isotropic oscillator-ghost-oscillator
system. It is straightforward to verify that the general solution to this
system may be expressed in terms of three unconstrained parameters
$\{A,\theta_1,\theta_2\}$ and belongs to the family of parametrised curves
$$t\mapsto\{X=A\,\cos(\omega t+\theta_1),\,Y=\eta A\,\cos(\omega
t+\theta_2)\}\Eqno$$
where $\eta=\pm 1$. For any domain for $t$ these  lie on
1-dimensional loci in ${\bf R}^2$ defined by
$$X^2+Y^2-2\eta \cos\theta XY -A^2\sin^2\theta =0\Eqno \label\orbits$$
where $\theta=\theta_2-\theta_1$.
These  are ellipses with major axes  oriented at $\pi/4$ to the
positive/negative X-axis according as $\eta$ is $\pm 1$, and eccentricity
and scale determined by $\theta$ and $A$ respectively.

The inverse of the mappings \mapa\ , \mapb\  follows from:
$$e^{-c\psi}=\sqrtc(x-y)\Eqno$$
$$\frac{1}{c\chi}=x^2-y^2.\Eqno$$
So for real $\psi$ we have $x\geq y$. This condition places the following
constraint on the parameters of the model:
$$\frac{\alpha\sqrtc}{\omega^2}\equiv
\frac{\alpha}{\vert\Lambda_0\vert\sqrtc}\geq 2
\vert A\cos\left ((1+\eta)\pf+\frac{(\theta_1-\theta_2)}{2}\right )\vert
.\Eqno$$
In theses circumstances
$$\frac{1}{\chi}=-c A^2 \sin(\thm)\sin(2\omega t+\thp)+$$
$$
2 \frac{A}{\omega^2}\alpha
c^{3/2}\eta\cos\left\{(1-\eta)\pf+
\frac{\thm}{2}\right\}\cos\left\{\omega
t+(1-\eta)\pf+\frac{\thm}{2}\right\}.\Eqno$$
Consider the special solution given by
$\eta=1,\,\,\thp=2\pi n,\,\, \thm=2\pi m$ where $m,\, n$ are integers:
$$\frac{1}{\chi}=2\,(-1)^{m+n} \frac{A}{\omega^2}\alpha c^{3/2}\cos(\omega
t)\Eqno$$
 $${\cal R}=2\,\chi \frac{d^2}{dt^2}\left(\frac{1}{\chi}\right)
=-2\, \omega^2=2\Lambda_0 c.\Eqno$$

We recognise here a solution that can be used to construct a degenerate
metric. For under the coordinate transformation $t=\frac{2}{3}\beta^{3/2}$
with $t>0$, $dt\otimes dt$ maps to $-\beta d\beta\otimes d\beta$ and
$\cos(\frac{2}{3}\omega\beta^{3/2})$ is a real function of $\beta$. Thus
one may construct a metric that changes from Euclidean to Lorentzian
signature across the hypersurface $\beta =0$.

\def\norm{\left (\frac{2\omega}{\pi}\right )^{1/2}}
\def\gauss{e^{-\frac{1}{2}(z_1^2+z_2^2)}}

\Section{Quantum States}

The classical cosmologies \orbits\ are the ``zero-energy'' solutions of the
Lagrangian \lag\ . The Hamiltonian for the theory is
$$ H=(\rp4 p_X^2+\omega^2 X^2)-(\rp4 p_Y^2+\omega^2 Y^2)\Eqno$$
so a basis of solutions of the Wheeler-DeWitt equation,
$$ H\Psi(X,Y)=0\Eqno \label\wdwitt$$
in a representation with $\hbar = 1,\,\,p_X=\frac{1}{i}\partial_X$ and
$p_Y=\frac{1}{i}\partial_Y$,
may be immediately expressed in terms
of  classical Hermite functions. A
general solution in terms of $z_1=\sqrt{2\omega} X,\,z_2=\sqrt{2\omega} Y$ is
$$\Psi(z_1,z_2)=\norm\sum^\infty_{n=0}\,\frac{c_n}{2^n n!}\,\gauss H_n(z_1)
\,H_n(z_2)\Eqno \label\wfn$$
 where $\{c_n\}$ is a set of complex constants and $H_n(z)$ is a Hermite
polynomial of order $n$.
For our subsequent analysis it will be very useful to construct a family of
particular solutions by choosing a particular set $\{c_n\}$.
 With the aid of the representation
\def\psq#1{\partial^2_{#1}}
$$H_n(z)=\left\{\exp\left (-\rp4\psq{z}\right )\right\}\,\,(2 z)^n\Eqno$$
we may express
$$\Psi(z_1,z_2)=B\norm\,\gauss \,e^{-\rp4 (\psq{z_1}+\psq{z_2})}\,
e^{\bf{Z^T\, S\, Z}}\Eqno$$
where ${\bf{Z^T}}=(z_1\,\,z_2)$,
$${\bf{S}}=\pmatrix{0 {\quad \,\,\,\,\tanh(\zeta/2)}\cr \tanh(\zeta/2){\quad
0}}\Eqno$$
and we choose $c_n=B \,\tanh^n(\zeta/2)$ where $\zeta,\,B$ are arbitrary
complex numbers. (This parametrisation for $c_n$  considerably simplifies
our resulting amplitude.)
With the aid of the formula \Ref{J D Louck, Adv. App. Math. {\bf 2} (1981) 239}
$$e^{-\rp4 (\psq{z_1}+\psq{z_2})}\,e^{\bf{Z^T\, S\, Z}}=
[det({\bf{I+S}})]^{-\rp2}\,e^{\bf{Z^T\,S(I+S)}^{-1}{\bf \,Z}}\Eqno$$
 we find that \wfn\ may be written

%$$\Psi(z_1,z_2)=\frac{B}{(1-\beta^2)^{\rp2}}\norm \exp\left\{-\btaz
%(z_1^2+z_2^2)+2\btaa z_1 z_2\right\}\Eqno$$
$$\Psi_{\mu,\nu}^\eta (X,Y)=A_0\,\exp\left\{-\omega\cosh\mu\cos\nu
[X^2+Y^2-2\eta\tanh\mu \,X Y]\right\}$$
$$\quad\quad\quad\quad\quad\quad \exp\left\{-i\omega\sinh\mu\sin\nu
[X^2+Y^2-2\eta\coth\mu \,X Y]\right\}\Eqno \label\state$$
where $\mu+i\nu = \zeta$ for real $\mu\,,\nu$, $\eta=\pm 1$ and $A_0$ is an
arbitrary complex constant. These
solutions
have the interesting property that for all $\mu\, ,\nu\,,\eta$ the modulus of
each
is constant on the family of ellipses
$$X^2+Y^2-2\eta\tanh\mu \,X Y=constant\Eqno$$  while the associated
phase is constant on the family of hyperbolae
$$X^2+Y^2-2\eta\coth\mu\, X Y=constant.\Eqno$$
It should be noted that the family of ellipses coincide with the classical
loci \orbits\ . The solution \state\ has good asymptotic behavior in the
$XY$ plane if $-\pi/2 < \nu < \pi/2 $.

The Wheeler-DeWitt equation \wdwitt\ endows ${\bf R}^2$ with a natural
(Lorentzian signature) metric ${\cal G}$. In terms of the coordinates
$\{X,Y\}$:
$${\cal G}=\partial X\otimes \partial X -\partial Y\otimes \partial
Y.\Eqno \label\metric$$
If $*$ denotes the associated Hodge map then \wdwitt\ may be written:
$$d*d\Psi-W*\Psi=0\Eqno \label\wdw$$
where $W(X,Y)=4\omega^2(X^2-Y^2)$. By multiplying \wdw\ by
$\bar\Psi$ and subtracting from the corresponding equation obtained by
complex conjugation we readily verify that
$$d\,{\cal J}=0\Eqno$$
where the current 1-form
$${\cal J}=Im(\bar\Psi *d\,\Psi).\Eqno$$
If we express the amplitude in polar form, $\Psi=R\exp(i S)$, then
$${\cal J}=R^2 d\,S.\Eqno$$ The gradient of the phase of $\Psi$ contains
information which can be used to construct wavepackets that have local
maxima in the
vicinity of classical loci. Indeed if we examine the vector field
$\widetilde{*{\cal J}}$ we find
$$\widetilde{*{\cal J}}=R^2\left\{(\partial_YS)\,\partial_Y
-(\partial_XS)\,\partial_X\right\}.\Eqno \label\vecfld$$
In these expressions the metric dual is taken with respect to the metric
\metric\ so
$\widetilde{d\,S}={\cal G}(d\, S,-)$. For the above amplitude \state\
 the integral
curves  of this vector field are solutions of the equation
$$\frac{dY}{dX}=-\frac{Y-\eta\coth\mu X}{X-\eta\coth\mu Y}\Eqno$$
with solutions
$$X^2+Y^2-2\eta\tanh\mu\, X Y =constant.\Eqno$$
Thus the integral curves coincide with the classical orbits \orbits\ .
This analysis mirrors the semi-classical description of
  point particles in terms
of the rays orthogonal to the wave fronts of  plane wave solutions in
non-relativistic quantum mechanics. There are however a number of
important differences. In particular the natural metric of the
configuration space on which the amplitudes \state\
are defined is not Riemannian.
This makes their interpretation in terms of a conserved non-negative
probability density difficult.
Nevertheless it is commonly believed
 \Ref{C Kiefer, Nucl. Phys. B341 (1990)  273} \label\kiefer
 that for any solution $\Phi$ of \wdw\
with dominant support in the vicinity of some classical loci,
the real quantity  $\vert\Phi\vert^2$ is related to some
relative measure of the occurrence of the corresponding
 classical configuration. It is of
interest therefore to see if a superposition of states of the form
 \state\  can be constructed that will yield ``elliptical wavepackets'' that
accentuate a set of the classical orbits \orbits\ .

Guided by the structure of the phase information contained in \state\ the
following superposition is suggested:
$$\Phi(X,Y)=\Psi^\eta_{\mu,\nu}(X,Y)-
\Psi^\eta_{\mu+\delta\mu,\nu+\delta\nu}(X,Y)\Eqno$$
 for some $\delta\mu,\delta\nu$ in the vicinity of $\mu,\nu$.
Figures 1 and 2 illustrate this exact solution for $\mu=1,\delta\mu=0.2,
\nu=\pi/4,\delta\nu=3\pi/20$ and $\eta=1,\,\, \omega=1$.
These profiles clearly  delineate a set of
the classical orbits \orbits\ .

The above discussion has assumed that $\omega^2 \geq 0$. If we continue
$\omega$ to $i\Omega$ with $\Omega$ real, then (with $U\neq 0$, $\Lambda_0
c=\Omega^2$)
 the classical theory corresponds to  the motion
of a two-dimensional constrained ``hyperbolic oscillator''. The classical
orbits are  hyperbolae and one readily verifies that the above discussion
can be applied to the Wheeler-DeWitt solutions
$$\Gamma_{\mu,\nu}^\eta (X,Y)=A_0\,\exp\left\{\Omega\sinh\mu\sin\nu
[X^2+Y^2-2\eta\coth\mu \,\,X Y]\right\}$$
$$\quad\quad\quad\quad  \exp\left\{-i\Omega\cosh\mu\cos\nu
[X^2+Y^2-2\eta\tanh\mu \,\,X Y]\right\}\Eqno \label\hypstates$$
with corresponding conclusions. Thus it is possible to superpose
neighbouring members of the set \hypstates\ to generate packets that delineate
a set of classical hyperbolic orbits. In this case no choice of $\mu$ or
$\nu$
exists  which makes \hypstates\ asymptotically well behaved everywhere.
Figures 3 and 4 illustrate the exact solution
$$\Phi(X,Y)=\Gamma^\eta_{\mu,\nu}(X,Y)+
\Gamma^\eta_{\mu+\delta\mu,\nu+\delta\nu}(X,Y)\Eqno$$
 for $\mu=1.317,\delta\mu=-1.217,
\nu=-0.0115,\delta\nu=0.002$ and $\eta=1,\,\, \Omega=1$.

\def\alphaa{a}

If we take $\Lambda_0=0$  but $\alpha=\alphaa /\sqrt{c}\neq 0$ in the potential
\pot\
then the classical equations of motion become

$$\ddot x=\frac{\alphaa}{2}\Eqno$$
$$\ddot y=-\frac{\alphaa}{2}\Eqno$$
with solutions constrained by
$$\dot x^2-\dot y^2 -\alphaa (x+y)=0.\Eqno$$
The classical orbits follow as the parabolas
$$(x+y)^2-\frac{2}{\alphaa}(b_1+b_2)^2(x-y)+c_1=0\Eqno \label\parab$$
where $c_1=(b_1+b_2)^2\{4\alphaa d_1 -(b_1-b_2)(3 b_1+b_2)\}/\alphaa^2$ with
$d_1,\,b_1$ and $b_2$
unconstrained real constants.
The Wheeler-DeWitt equation for this system is \wdw\ with $W=-\alphaa (x+y)$.
Its separable solutions  may be expressed in terms of Airy functions. The
particular solution of relevance to our discussion is
$$\Theta_\beta(x,y)=A_0
\exp\left\{-\frac{\alphaa(x+y)^2}{2\beta}+\alphaa\beta(x-y)\right\}\Eqno
\label\wfnn$$
where $\beta$ is an arbitrary complex constant. With $\mu+i\nu =\beta$ this
becomes
$$\Theta_{\mu\,,\nu}(x,y)
=A_0 \exp\left\{\frac{-\alphaa\mu}{2(\mu^2+\nu^2)}[(x+y)^2
                 -\frac{2(\mu^2+\nu^2)}{\alphaa}(x-y)]\right\}$$
$$\quad\quad\quad\quad\quad
\exp\left\{\frac{i\alphaa\nu}{2(\mu^2+\nu^2)}[(x+y)^2
                 +\frac{2(\mu^2+\nu^2)}{\alphaa}(x-y)]\right\}\Eqno
\label\wfnnn$$

Again we observe that for all $\mu\,,\nu$ the modulus of \wfnnn\ is constant
on the loci \parab\ . The integral curves of the vector field
\vecfld\  computed from the phase of the wavefunction \wfnnn\ are solutions
of
$$\frac{dy}{dx}=-\frac{\alphaa (x+y)-\mu^2-\nu^2}{\alphaa
(x+y)+\mu^2+\nu^2}\Eqno $$ and coincide with the classical orbits \parab\
. Although no choice of $\mu\,,\nu$ exists which makes \wfnnn\
asymptotically well behaved everywhere in the $xy$ plane it is possible to
construct solutions that delineate a set of classical parabolic orbits.
Figures  5 and 6 illustrate the solution
$$\Phi(x,y)=\Theta_{\mu,\nu}(x,y)-
\Theta_{\mu+\delta\mu,\nu+\delta\nu}(x,y)\Eqno$$
for $\mu=0.01,\delta\mu=0,
\nu=1,\delta\nu=0.02\,\,, \alphaa=1$.

 For the special case
$U=0$ ($\omega^2=0$)  the classical orbits are  lines parallel to the
lines $x=\pm y $,
corresponding to flat space-time geometries.
Then \wdw\ degenerates into the two-dimensional wave equation with the well
known D'Alambertian solution in terms of two arbitrary functions of $x\pm y$.

%\vfill
%\eject

\Section{Discussion}

Within the framework of this two-dimensional field theory we have drawn
attention to an interesting property of a family of Wheeler-DeWitt
solutions. These solutions are associated with a class of potentials that
are expressible in terms of a general quadratic form in a set of variables
that linearises the classical field equations for homogeneous geometries.
In terms of these variables the classical solutions may be represented as
conic sections in a two-dimensional configuration space. Particular
Wheeler-DeWitt solutions are expressible in terms of a complex parameter so
that their phase and modulus are correlated in an interesting way. The
integral curves of the vector field associated, via a flat Lorentzian metric,
with a closed form generated by such solutions coincide exactly with these
classical conic sections.
This phenomena may be contrasted with the relation of classical Newtonian
particle trajectories to the integral curves of the vector field associated
with the conserved probability current in non-relativistic quantum
mechanics.
Despite the absence of a preferred time variable in quantum gravity and a
Euclidean signature metric on configuration space we have demonstrated that
it is possible to find a precise correlation between classical solutions
and particular quantum states. Configuration space is foliated by families
of curves along which the phase and modulus of these states are constant.
Furthermore these families of curves are orthogonal with respect to the
Lorentzian metric \metric\ .
We have also demonstrated that these states are effective in constructing
non-dispersive
wavepackets that accentuate a class of classical configurations.
It would be of interest to know if these
properties  are specific to the models considered here.

We believe the above results  offer an insight into the relation between
classical and quantum cosmologies and that the particular solutions
that we have constructed  may serve as a useful starting point
for a more ambitious investigation of the role played by the constraints
in the description of non-homogeneous geometries.

\bigskip\leftline{\bf ACKNOWLEDGMENTS}\nobreak

This research was supported in part by the European Union under the Human
Capital and Mobility programme.
M \"O
 wishes to thank the School of Physics and Materials, University of
Lancaster for its hospitality and the Scientific and Technical Research
Council of Turkey (T\"UB${\dot {\hbox{I}}}$TAK) for support.

\vfill
\eject

{\centerline{\bf Figure Captions}}
\vskip 1cm

\centerline{Figure 1}
\noindent Surface plot of the Wheeler-DeWitt Solution
$\Phi(X,Y)=\Psi^\eta_{\mu,\nu}(X,Y)-
\Psi^\eta_{\mu+\delta\mu,\nu+\delta\nu}(X,Y)$
with
 $\mu=1,\delta\mu=0.2,
\nu=\pi/4,\delta\nu=3\pi/20$ and $\eta=1,\,\, \omega=1$.
\vskip 0.5cm
\centerline{Figure 2}
\noindent Plot of the Wheeler-DeWitt Solution
$\Phi(X,Y)=\Psi^\eta_{\mu,\nu}(X,Y)-
\Psi^\eta_{\mu+\delta\mu,\nu+\delta\nu}(X,Y)$
with
 $\mu=1,\delta\mu=0.2,
\nu=\pi/4,\delta\nu=3\pi/20$ and $\eta=1,\,\, \omega=1$
displaying elliptical classical orbits.
\vskip 0.5cm
\centerline{Figure 3}
\noindent Surface plot of the Wheeler-DeWitt Solution
$\Phi(X,Y)=\Gamma^\eta_{\mu,\nu}(X,Y)+
\Gamma^\eta_{\mu+\delta\mu,\nu+\delta\nu}(X,Y)$
with $\mu=1.317,\delta\mu=-1.217,
\nu=-0.0115,\delta\nu=0.002$ and $\eta=1,\,\, \Omega=1$.
\vskip 0.5cm
\centerline{Figure 4}
\noindent Plot of the Wheeler-DeWitt Solution
$\Phi(X,Y)=\Gamma^\eta_{\mu,\nu}(X,Y)+
\Gamma^\eta_{\mu+\delta\mu,\nu+\delta\nu}(X,Y)$
with $\mu=1.317,\delta\mu=-1.217,
\nu=-0.0115,\delta\nu=0.002$ and $\eta=1,\,\, \Omega=1$
displaying hyperbolic classical orbits.
\vskip 0.5cm
\centerline{Figure 5}
\noindent Surface plot of the Wheeler-DeWitt Solution
 $\Phi(x,y)=\Theta_{\mu,\nu}(x,y)-
\Theta_{\mu+\delta\mu,\nu+\delta\nu}(x,y)$
with $\mu=0.01,\delta\mu=0,
\nu=1,\delta\nu=0.02\,\,, \alphaa=1$.
\vskip 0.5cm
\centerline{Figure 6}
\noindent Plot of the Wheeler-DeWitt Solution
 $\Phi(x,y)=\Theta_{\mu,\nu}(x,y)-
\Theta_{\mu+\delta\mu,\nu+\delta\nu}(x,y)$
with $\mu=0.01,\delta\mu=0,
\nu=1,\delta\nu=0.02\,\,, \alphaa=1$
displaying parabolic classical orbits.

\vfill
\eject

\References

\bye